\documentclass{aa} 
\usepackage{epsfig}
\def\LB{$\lambda$\,Bootis } 
\begin{document} 
\title{On the Period-Luminosity-Colour-Metallicity relation
and the pulsational characteristics of \LB type stars\thanks{Based on observations 
from the Austrian Automatic
Photoelectric Telescope (Fairborn Observatory), SAAO and Siding Spring Observatory}} 
\author{E.~Paunzen\inst{1,2}, G.~Handler\inst{3}, W.W.~Weiss\inst{1},
N.~Nesvacil\inst{1}, A.~Hempel\inst{4,5}, E.~Romero-Colmenero\inst{3}, 
F.F.~Vuthela\inst{3,6}, P.~Reegen\inst{1}, R.R.~Shobbrook\inst{7}, D.~Kilkenny\inst{3}} 
\offprints{E.~Paunzen}
\mail{Ernst.Paunzen@univie.ac.at}
\institute{Institut f\"ur Astronomie der Universit\"at Wien,
        T\"urkenschanzstr. 17, 1180 Wien, Austria
\and
        Zentraler Informatikdienst der Universit\"at Wien,
		Universit\"atsstr. 7, 1010 Wien, Austria
\and
		South African Astronomical Observatory, P.O. Box 9, Observatory 7935, 
		South Africa
\and	
		Department of Physics, University of Cape Town, Private
		Bag, Rondebosch 7701, South Africa
\and
		Max-Planck-Institut f\"ur Astronomie,
		K\"onigsstuhl 17, 69117 Heidelberg, Germany
\and
		Department of Physics, University of the North-West,
		Private Bag X2046, Mmabatho 2735, South Africa
\and	
		Research School of Astronomy and Astrophysics, 
		Australian National University, Canberra, ACT 0200, Australia
		}
\date{Received 2002; accepted 2002}
\authorrunning{Paunzen, Handler, et al.}
\titlerunning{PLC-relation for \LB type stars}
\abstract{
Generally, chemical peculiarity found for stars on the upper main
sequence excludes $\delta$ Scuti type pulsation (e.g. Ap and Am stars), but
for the group of \LB stars it is just the opposite. This makes them very
interesting for asteroseismological investigations. The group of \LB type
stars comprises late B- to early F-type, Population\,I objects which
are basically metal weak, in particular the Fe group elements, but
with the clear exception of C, N, O and S.
The present work is a continuation of the studies by Paunzen et al. (1997,
1998), who presented first results on the pulsational characteristics of
the \LB stars. Since then, we have observed 22 additional objects; we
found eight new pulsators and confirmed another one. Furthermore, new
spectroscopic data (Paunzen 2001) allowed us to sort out misidentified
candidates and to add true members to the group. From 67 members
of this group, only two are not photometrically investigated yet which
makes our analysis highly representative.
We have compared our results on the pulsational behaviour of the \LB stars
with those of a sample of $\delta$ Scuti type objects. We find that at
least 70\% of all \LB type stars inside the classical instability strip
pulsate, and they do so with high overtone modes ($Q$\,$<$\,0.020\,d). Only
a few stars, if any, pulsate in the fundamental mode. Our photometric
results are in excellent agreement with the spectroscopic work on
high-degree nonradial pulsations by Bohlender et al. (1999). Compared to
the $\delta$ Scuti stars, the cool and hot borders of the instability
strip of the \LB stars are shifted by about 25\,mmag,
towards smaller $(b-y)_0$.
Using published abundances and the metallicity sensitive indices of the
Geneva 7-colour and Str\"omgren $uvby\beta$ systems, we have derived [Z]
values which describe the surface abundance of the heavier elements for
the group members. We find that the 
Period-Luminosity-Colour relation for the group of \LB stars is within the errors
identical with that of the normal $\delta$ Scuti stars. No clear evidence
for a statistically significant metallicity term was detected.
\keywords{Stars -- \LB; stars -- chemically peculiar; stars -- early type} 
}
\maketitle

\section{Introduction} 

In this paper we present an extensive survey to analyse the pulsational
characteristics of the \LB stars. This small group comprises late B- to
early F-type, Population\,I stars which are metal weak (particularly the
Fe group elements), but with the clear exception of C, N, O and S. Only a
maximum of about 2\% of all objects in the relevant spectral domain are
believed to be \LB type stars.

Several theories were developed to explain the peculiar abundance pattern
for members of this group. The most acknowledged models include diffusion
as main mechanism together either with mass-loss (Michaud \& Charland
1986, Charbonneau 1993) or with accretion of circumstellar material (Venn
\& Lambert 1990, Turcotte \& Charbonneau 1993). Another two theories deal
with the influence of binarity on this phenomenon (Andrievsky 1997,
Faraggiana \& Bonifacio 1999). Heiter (2002) and Heiter et. al (2002) also
tried to explain the abundance pattern in the context of the proposed
theories.

In general, chemical peculiarity inhibits $\delta$ Scuti type pulsation
(e.g. for Ap and Am stars, see Kurtz 2000 for a recent discussion) but for the
group of \LB stars it is just the opposite. In two previous studies (Paunzen et
al. 1997, 1998), we presented non-variable as well as pulsating \LB
stars. Since then, we have observed 22 additional objects and found eight new
pulsators and confirmed another. Furthermore, new spectroscopic data (Paunzen
2001) has allowed us to sort out misidentified candidates and to add true
members of the group.

Turcotte et al. (2000) investigated the effect of diffusion (probably the
main cause of the \LB phenomenon) on the pulsation of stars at the upper main
sequence. Although these authors mainly investigated the theoretical
behaviour of apparently metal-rich objects, their conclusions also have an
impact for the \LB group: little direct pulsational excitation from
Fe-peak elements was found, but effects due to settling of helium along
with the enhancement of hydrogen are important. Turcotte et al. (2000)
find that, as their models of peculiar stars evolve, they become generally
pulsational unstable near the red edge of the instability strip, whereas
the behaviour at the blue edge is mainly sensitive to the surface metal
abundance. Although the proposed models are still simplified (e.g.
treatment of convection) these preliminary results already point towards the
most important effects on the theoretical pulsational instability and
behaviour of chemically peculiar stars.

The aim of the present paper is to analyse the pulsational characteristics
of the group of \LB stars and to test for the presence of a possible
Period-Luminosity-Colour-Metallicity relation. The latter is especially
interesting in the light of the models by Turcotte et al. (2000). The
pulsational characteristics of the \LB group (e.g. ratio of variable to
non-variable objects and distribution of pulsational constants) may help
to put tighter constraints on these models.

\begin{table}
\caption[]{Sites, dates and telescopes used for our survey}
\label{sites}
\begin{center}
\begin{tabular}{lcccc}
\hline
Site & Date & Telescope & Stars & Ref. \\
\hline
APT (Fairborn) & 05.2001 & 0.75 & 3 & 1 \\
SAAO & 04.2001 & 0.50 & 8 & 2 \\
     & 07.2001 \\
	 & 08.2001 \\
	 & 09.2001 \\
	 & 10.2001 \\
SAAO & 12.2000 & 0.75 & 9 & 3 \\
	 & 01.2001 \\
SAAO & 08.2001 & 1.00 & 1 & 4 \\
Siding Spring & 01.2002 & 0.60 & 1 & 5 \\
\hline
\end{tabular}
\end{center}
\end{table}

\begin{table*}
\caption[]{Observing log of eight newly discovered and one confirmed
(HD~75654) pulsating \LB stars. Some information on the comparison stars
is also given. The differential light curves are shown in
Fig. \ref{lightcurve}.}
\label{log1}
\begin{center}
\begin{tabular}{rccccccc}
\hline
HD & JD & hrs & $m_{\rm V}$ & Spec. & Freq. & Amp. & Ref. \\
& & & [mag] & & [d$^{-1}$] & [mag] \\
\hline
13755 & 2451899 & 3.2 & 7.84 & $\lambda$\,Boo & 12.50 & 0.015 & 3 \\
	  & 2451903 & 2.5 &		 & 	   & 16.85 & 0.007 \\
	  & 2451905 & 3.1 \\
	  & 2451909 & 3.1 \\
13602 &			& 	  & 8.52 & F6 \\
13710 &			&	  & 8.32 & K5 \\
\hline
35242 & 2451900 & 2.1 & 6.35 & $\lambda$\,Boo & 38.61 & 0.005 & 3 \\
	  & 2451902 & 5.4 &		 & 	   & 34.16 & 0.003 \\
	  & 2451908 & 3.3 &		 & 	   & 41.33 & 0.003 \\
35134 &			& 	  &	6.74 & A0 \\
34888 &			&	  & 6.78 & A5 \\
\hline
42503 & 2452291 & 4.2 & 7.45 & $\lambda$\,Boo & 7.00 & 0.015 & 5 \\
      & 2452292 & 1.9 \\
42058 & 		& 	  & 6.99 & A0 \\
43452 &		    &	  & 7.71 & F5 \\
\hline
75654 & 2451898 & 3.0 & 6.38 & $\lambda$\,Boo & 14.80 & 0.005 & 3 \\
	  & 2451902 & 1.8 &		 & 	   & 15.99 & 0.002 \\
	  & 2451905 & 3.1 \\
	  & 2451906 & 1.2 \\
	  & 2451907 & 3.8 \\
	  & 2451909 & 3.6 \\
74978 &			& 	  & 6.87 & A1 \\
75272 &			&	  & 6.98 & B9.5 \\
\hline
111604 & 2452061 & 4.1 & 5.89 & $\lambda$\,Boo & 8.77 & 0.020 & 1 \\
112412 & 		 & 	   & 5.61 & F1 \\
110375 &		 &	   & 8.33 & F5 \\
\hline
120896 & 2452097 & 3.9 & 8.50 & $\lambda$\,Boo & 17.79 & 0.010 & 2 \\
121372 & 		 & 	   & 8.67 & G5 \\
\hline
148638 & 2452097 & 4.6 & 7.90 & $\lambda$\,Boo & 16.32 & 0.016 & 2 \\
	   & 2452123 & 5.0 \\
148596 & 		 & 	   & 8.60 & F2 \\
148573 &		 &	   & 8.63 & B9 \\
\hline
213669 & 2451823 & 6.5 & 7.42 & $\lambda$\,Boo & 15.01 & 0.023 & 2 \\
       & 2451826 & 1.6 \\
	   & 2451827 & 1.1 \\
211878 & 		 & 	   & 7.70 & F5 \\
214390 & 		 & 	   & 7.90 & F3 \\
\hline 
290799 & 2451904 & 3.0 & 10.63 & $\lambda$\,Boo & 23.53 & 0.006 & 3 \\
	   & 2451906 & 4.8 \\
37652  & 		 & 	   & 7.35 & F5 \\
290798 &		 &	   & 10.40 & A2 \\
\hline
\end{tabular}
\end{center}
\end{table*}

\section{Program stars, observations and reductions}

Since our previous works (Paunzen et al. 1997, 1998) several
then-selected group members were investigated with classification resolution
spectroscopy and found to be misclassified. These are: HD~66920, HD~79025,
HD~82573, HD~141851, HD~143148, HD~145782, HD~149303, HD~179791, HD~188164 and
HD~192424 (Paunzen 2001). In total, 65 members were selected 
from the lists of Gray \& Corbally (1993) and
Paunzen (2001) which contain well established as well as good candidate \LB
type objects. Together with the
two newly discovered objects (HD~42503 and HD~213669; Sect. \ref{new_lbs}), we
have a sample of 67 \LB type stars.

\begin{figure*}
\begin{center}
\epsfxsize = 164mm
\epsffile{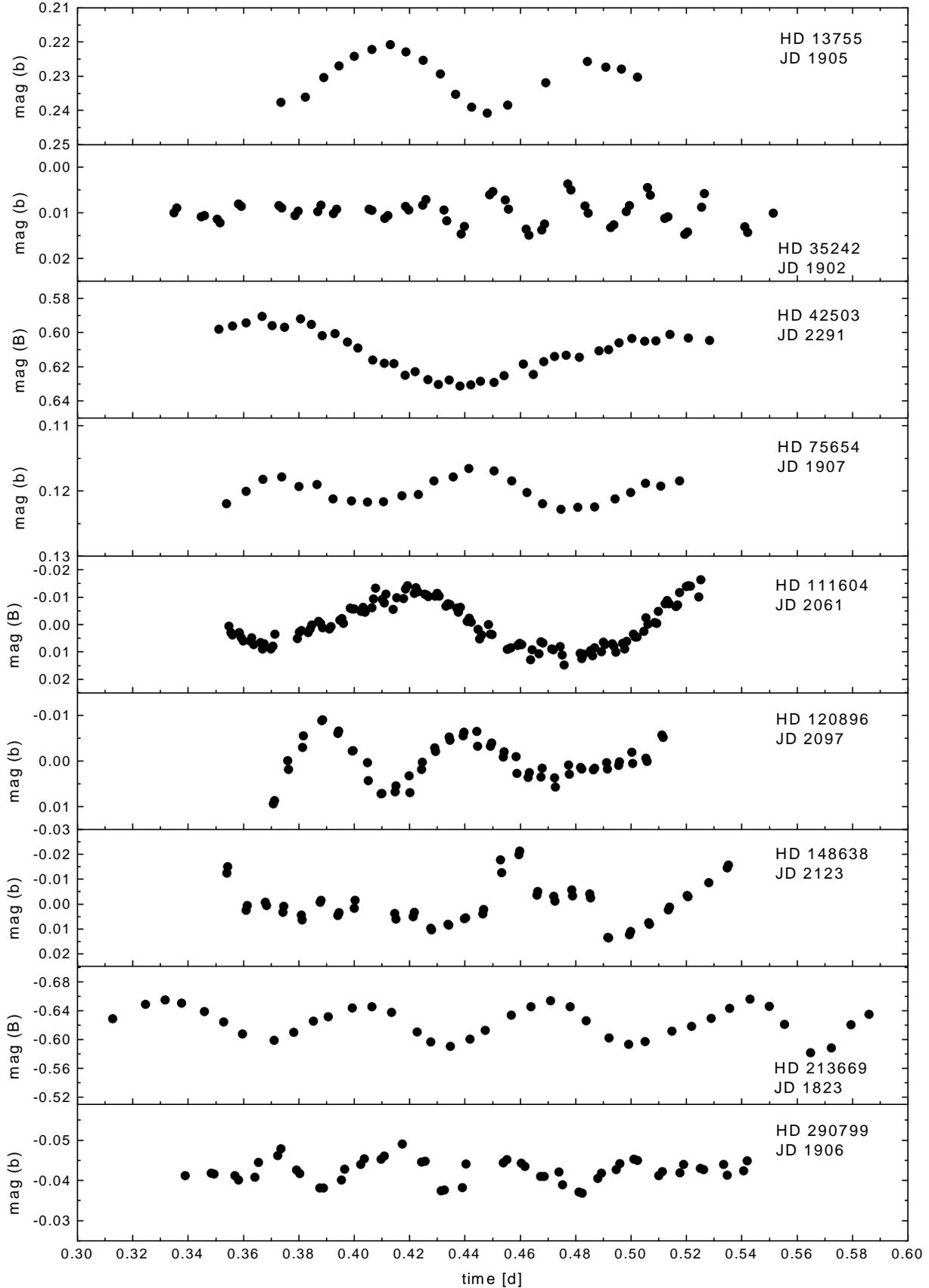}
\caption{Differential light curves of eight newly discovered and one
confirmed (HD~75654) pulsating \LB stars in Str\"omgren $b$ and
Johnson $B$; the dates of the
corresponding nights are given as JD 2450000+ 
(Table \ref{log1}).} \label{lightcurve}
\end{center}
\end{figure*}

The photometric observations were performed as described by Paunzen et al.
(1998) using photoelectric detectors (except for Ref. ``4'', Table
\ref{sites}, for which a CCD was used) and (if possible) two comparison
stars. A standard reduction method for dealing with dead-time, dark counts
and tube drifts was applied. The sky measurements (typically one per half
hour) were subtracted and differential light curves were generated.  For
the reduction of the CCD frames for HD~290492 the standard SAAO reduction
package as well as the program MOMF (Kjeldsen \& Frandsen 1992) were used.
Figure \ref{lightcurve} shows light curves of some of our variable
program stars.

Frequencies and amplitudes for the variable program stars (listed in Table
\ref{log1}) were derived using a standard Fourier algorithm (Deeming
1975). An analysis with the Phase-Dispersion-Minimization (Stellingwerf
1978) gave essentially the same results. A star is considered to be constant,
if the Fourier spectrum of the differential light curve does not contain
a statistically significant peak (Paunzen et al. 1997). These objects are listed in
Table \ref{log2}. 

\subsection{Previously known pulsating \LB stars}

The following nine stars were already known as variable. With the only
exception of HD~75654 (see below), they have not been re-observed by us:

\begin{itemize}
\item HD 6870: Breger (1979) lists a period of 94 minutes and an amplitude of
15\,mmag for Johnson [V]. 
\item HD~11413: This star is multiperiodic (Waelkens \& Rufener 1983) with a dominant
frequency of 54 minutes and an amplitude of 18\,mmag. New measurements (Koen, private
communication) confirm the multiperiodic pulsations of this object.
\item HD~15165: The multiperiodicity of VW Arietis motivated
the fifth STEPHI campaign (Liu et al. 1996).
Seven significant periods between 1.8 and 3.9 hr were found in a data set
of about 150 hr. 
\item HD~75654: This object was discovered as a $\delta$ Scuti type pulsator by
Balona (1977), who reported a period of 0.087\,d (11.49\,d$^{-1}$). Since no
other photometric measurements were published, we decided to re-observe
this object. We find two frequencies (14.80 and 15.99\,d$^{-1}$) based on
observations during six nights.
\item HD~87271: The suspected variability of this star (Handler 2002) was confirmed by
Handler et al. (2000). Although the measured light curve spans only several hours,
multiperiodicity with a time scale of about 80 minutes is evident. 
\item HD~105759: Martinez et al. (1998) performed a multisite
campaign, detecting five pulsation periods between 1.0 and 2.8 hr, as well
as a detailed abundance analysis for this star. 
\item HD~110377: Radial velocity variations as well as multiperiodicity
with periods between 0.5 and 2.0 hr were reported by Bartolini et
al. (1980b). Evidence for amplitude and frequency variations makes this
object very interesting for detailed follow-up investigations. Such
observations are however beyond the scope of this paper. 
\item HD~153747: Desikachary \& McInally (1979) reported multiperiodicity
(periods between 0.96 and 1.2 hr) as
well as a variable frequency spectrum of this object. Unfortunately no
further references on the pulsational behaviour of this star were found. 
\item HD~192640: This star's variability was discovered by Gies \& Percy
(1977). Since 1995 permanent multisite observations have been performed.
No detailed overall analysis has been published to date. Data subsets
(Kusakin \& Mkrtichian 1996; Paunzen \& Handler 1996; Mkrtichian et
al. 2000) suggest multiperiodicity with a main period of 38 minutes and an
amplitude of 20\,mmag in Johnson V.
\end{itemize}
We have used the published frequencies and amplitudes of these objects
from the above-mentioned references in our analysis. If more than one
frequency has been published, we have weighted the individual periods
with the squared amplitude to obtain a mean period.

\begin{table}
\caption[]{Observing log of thirteen \LB stars not found to pulsate as
well as some comparison star information.}
\label{log2}
\begin{center}
\begin{tabular}{rcccccc}
\hline
HD/BD & JD & hrs & $m_{\rm V}$ & Spec. & Limit & Ref \\
& & & [mag] & & [mmag] \\
\hline
7908 & 2451898 & 2.0 & 7.29 & $\lambda$\,Boo & 0.3 & 3 \\
     & 2451907 & 2.3 \\
7629 & 		   & 	 & 7.13 & A9 \\
7896 & 		   & 	 & 7.95 & G6 \\
\hline
24472 & 2451900 & 2.9 & 7.09 & $\lambda$\,Boo & 0.8 & 3 \\
24616 & 		& 	  & 6.70 & G8 \\
25385 & 		&	  & 7.40 & F0 \\
\hline
54272 & 2451908 & 2.6 & 8.80 & $\lambda$\,Boo & 1.4 & 3\\
54692 &			& 	  & 8.51 & A0 \\
+19\,622 & 		&	  & 8.90 & A2 \\
\hline
74873 & 2451904 & 3.5 & 5.89 & $\lambda$\,Boo & 1.6 & 3 \\
74228 &			& 	  & 5.65 & A3 \\
75108 & 		&	  & 8.38 & G5 \\
\hline
83277 & 2451901 & 3.5 & 8.30 & $\lambda$\,Boo & 1.4 & 3 \\
83547 & 		& 	  & 8.62 & A0 \\
82709 & 		&	  &	8.04 & A9 \\
\hline
90821 & 2452039 & 2.0 & 9.47 & $\lambda$\,Boo & 2.2 & 1 \\
90878 & 		& 	  &	7.82 & F8 \\
90748 & 		&	  & 8.67 & F8 \\
\hline
107223 & 2452003 & 5.2 & 7.35 & $\lambda$\,Boo & 1.9 & 2 \\
107143 & 		 & 	   & 7.87 & A1 \\
107265 & 		 &	   & 8.76 & A0 \\
\hline
111005 & 2452004 & 1.9 & 7.96 & $\lambda$\,Boo & 2.1 & 2\\
110705 & 		 & 	   & 8.36 & F0 \\
110989 &		 &	   & 8.41 & F8 \\
\hline
130767 & 2452039 & 5.5 & 6.91 & $\lambda$\,Boo & 1.2 & 1 \\
130556 & 		 & 	   & 7.84 & F1 \\
130396 &		 &	   & 7.41 & F8 \\
\hline
149130 & 2452127 & 5.6 & 8.50 & $\lambda$\,Boo & 2.4 & 2 \\
148597 & 		 & 	   & 8.25 & B9 \\
149471 & 		 &	   & 8.94 & F6 \\
\hline
216847 & 2452190 & 2.9 & 7.06 & $\lambda$\,Boo & 1.7 & 2 \\
	   & 2452191 & 3.2 \\
216349 & 		 & 	   & 7.84 & K1 \\
217686 & 		 &	   & 7.56 & F7 \\
\hline
290492 & 2451901 & 3.0 & 9.27 & $\lambda$\,Boo & 1.8 & 2 \\
	   & 2452190 & 2.2 & & & & \\
	   & 2452192 & 4.0 \\
290575 & 		 & 	   & 9.85 & F5 \\
$-$00\,984 &	 &	   & 8.37 & HgMn \\
\hline
261904 & 2452190 & 2.1 & 10.20 & $\lambda$\,Boo & 3.5 & 4 \\
	   & 2452191 & 2.2 \\
261941 & 		 & 	   & 10.94 & A2 \\
\hline
\end{tabular}
\end{center}
\end{table}

\subsection{Group members not observed by us}

Five well established members of the \LB group were not photometrically
investigated: HD~110411, HD~125889, HD~170680, HD~184779 and HD~198160. For
three of them (HD~110411, HD~170680 and HD~198160) photometric measurements in
the HIPPARCOS and TYCHO catalogues (ESA 1997) were found.  Since these
observations are not optimal to find $\delta$ Scuti type pulsation, only a
rough estimate for variability can be made. We find a level of non-variability
based on the HIPPARCOS photometry of 3\,mmag for HD~110411 and HD~170680, and
4\,mmag for HD~198160.  In fact, HD~110411 was suspected as variable by
Bartolini et al. (1980a), but Antonello \& Mantegazza (1982) concluded that
there is no evidence for periodic terms in the light curve: different
oscillation modes may be excited occasionally and then be damped again. We
therefore treat this star as being constant within a limit of 3\,mmag. 
Consequently, the other two objects (HD~125889 and HD~184779) were not
considered in the following analysis.

\subsection{Notes on individual stars}

In the following sections we describe special properties of some
individual stars in more detail.

\subsubsection{HD~42503 and HD~213669} \label{new_lbs}

These two objects were suspected $\delta$ Scuti type pulsators based on
HIPPARCOS photometry (Handler 2002).  Handler (1999) presented Str\"omgren
$uvby\beta$ photometry which puts these stars well within the typical area of
the \LB objects in a $m_{\rm 1}$ versus $(b-y)$ diagram 
(Paunzen et al.  1998). Our photometric measurements confirmed the
pulsation.

\begin{figure}
\begin{center}
\epsfxsize = 82mm
\epsffile{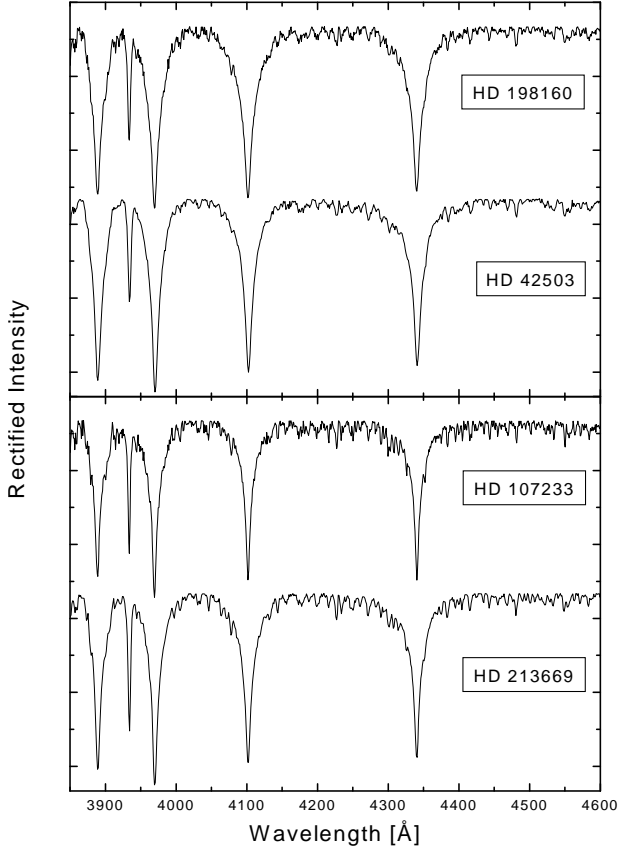}
\caption{Classification resolution spectra of the two newly discovered
\LB candidates HD~42503 (A2\,V\,$\lambda$ Boo;
upper panel) and HD~213669 (kA1hF0mA1\,V\,$\lambda$ Boo; lower panel)
together with the two well established objects HD~107233 (kA1hF0mA1\,V\,$\lambda$ Boo) and
HD~198160 (A2\,Vann\,$\lambda$ Boo) for comparison.} 
\label{spectra}
\end{center}
\end{figure}

We have performed additional spectroscopic observations to establish the nature
of these stars. These observations were done on the 1.9\,m telescope at SAAO in
the night of 03./04.10.2000. The Grating Spectrograph with the SITe CCD
together with the 600\,lines\,mm$^{-1}$ grating resulted in a useful wavelength
range of 1600\,\AA, a resolution of 2\,\AA\,and a signal-to-noise ratio of
about 200. The wavelength calibration was done with the help of a CuAr lamp
within standard IRAF routines. The classification was done within the system
described by Paunzen (2001). Both objects are very good \LB candidates, with
derived spectral types of A2\,V\,($\lambda$\,Boo) and kA1hF0mA1\,V\,$\lambda$
Boo for HD~42503 and HD~213669, respectively. 
The notation of the spectral classification is according to Gray (1988) where
$k$ stands for the classification of the \ion{Ca}{ii K} line, $h$ for the hydrogen
lines and $m$ for the appearance of the metallic-line spectrum compared to MK standards.
We therefore included them in our
sample. Figure \ref{spectra} shows their spectra together with those of two
well established \LB stars (HD~107233 and HD~198160; taken from Paunzen 2001)
which exhibit similar spectral characteristics. However, we note that a final
decision on their group membership has to be made after a detailed
determination of their chemical abundances (especially of C, N, O and
S) which was not done so far.

\subsubsection{HD~64491 and HD~111786}

Both stars are $\delta$ Scuti type pulsators and were reported as well
established members of the group by Gray (1988) and Paunzen \& Gray
(1997). However, both objects are spectroscopic binary systems (Faraggiana \&
Bonifacio 1999; Iliev et al. 2001). The published information on these stars
does not allow us to decide whether they are true \LB type objects, thus we
have not included either of them in our sample.

\subsubsection{HD~74873}

We have re-observed this object because the upper limit for non-variability
given by Paunzen et al. (1997) was very high (9.4\,mmag).  Our new observations
showed no variability within a limit of 1.6\,mmag.

\subsubsection{HD~175445}

This object exhibits a peculiar behaviour which we have, so far, been unable to
understand.  During our first short observing run (on 06/07 July 2001), its
magnitude and colour were consistent with their standard values from the literature
($b$\,=\,7.85, $v-b$\,=\,0.222), and
marginal evidence for pulsational variability was found. However, when we
attempted to re-observe the star on 05/06 August 2001, it appeared much fainter
and redder than in the previous month.

Consequently, we double-checked its correct identification and re-examined the
literature. We found no evidence for a misidentification or previous peculiar
behaviour. We obtained a 3\,hr light curve on the night of 06/07 August 2001,
where the object appeared as a star of $b$\,=\,10.67 and $v-b$\,=\,1.17. These
observations are consistent with an eclipse by a late K subgiant
companion. However, the evolutionary history of such a system would require the
binary to be close, and no other possible eclipses have yet been reported.

\subsubsection{HD~290799}

Paunzen et al. (1998) reported this star as constant with an upper limit of
10\,mmag in Geneva $V_{1}$. Our new observations show the star to be pulsating
with a frequency of 23.53\,d$^{-1}$ and an amplitude of 6\,mmag in Str\"omgren
$b$.

\begin{table*}
\caption
{Pulsating \LB stars; an asterisk denotes stars without an accurate
HIPPARCOS parallax; $\sigma$($b-y$)\,=\,$\pm$0.005\,mag. In
parenthesis are the errors in the final digits of the corresponding quantity.}
\label{pulsating}
\begin{center}
\begin{tabular}{lcllcllllll}
\hline
\multicolumn{1}{c}{HD} & $(b-y)_{0}$ & \multicolumn{1}{c}{log\,$T_{\rm eff}\,$} &
\multicolumn{1}{c}{log\,$g$} & \multicolumn{1}{c}{$v$\,sin\,$i$} &  
\multicolumn{1}{c}{[Z]} & \multicolumn{1}{c}{$M_{\rm V}$} &
\multicolumn{1}{c}{$M_{\rm B}$}  & \multicolumn{1}{c}{log\,$L_{\ast}/L_{\odot}$} &
\multicolumn{1}{c}{log\,$P$} & \multicolumn{1}{c}{$Q$} \\
& [mag] & \multicolumn{1}{c}{[dex]} & \multicolumn{1}{c}{[dex]} &  
\multicolumn{1}{c}{[km\,s$^{-1}$]} & \multicolumn{1}{c}{[dex]} & \multicolumn{1}{c}{[mag]} & 
\multicolumn{1}{c}{[mag]} & \multicolumn{1}{c}{[dex]} &
& \multicolumn{1}{c}{[d]}\\
\hline
6870	&	0.164	&	3.865(6)	&	3.84(11)	&	165	&	$-$1.03(20)	&	+2.29(42)	&	+2.20	&	1.02(17)	&	$-$1.19	&	0.023	\\
11413	&	0.104	&	3.899(7)	&	3.91(21)	&	125	&	$-$1.17(10)	&	+1.49(10)	&	+1.36	&	1.35(4)	&	$-$1.38	&	0.014	\\
13755	&	0.181	&	3.850(10)	&	3.26(10)	&	--	&	$-$0.75(30)	&	+0.93(10)	&	+0.83	&	1.57(4)	&	$-$1.12	&	0.010	\\
15165	&	0.189	&	3.846(12)	&	3.23(10)	&	90	&	$-$1.15(17)	&	+1.12(16)	&	+1.01	&	1.50(6)	&	$-$0.87	&	0.017	\\
30422	&	0.098	&	3.896(6)	&	4.00(20)	&	135	&	$-$1.50(20)	&	+2.35(2)	&	+2.23	&	1.01(1)	&	$-$1.68	&	0.010	\\
35242	&	0.058	&	3.916(5)	&	3.90(14)	&	90	&	$-$1.40(20)	&	+1.75(22)	&	+1.60	&	1.26(9)	&	$-$1.58	&	0.010	\\
42503	&	0.110	&	3.885(16)	&	3.10(10)	&	--	&	$-$0.83(20)	&	$-$0.03(4)	&	$-$0.14	&	1.96(2)	&	$-$0.85	&	0.013	\\
75654	&	0.158	&	3.866(6)	&	3.77(11)	&	45	&	$-$0.91(11)	&	+1.83(12)	&	+1.74	&	1.20(5)	&	$-$1.18	&	0.019	\\
83041	&	0.185	&	3.852(13)	&	3.76(20)	&	95	&	$-$1.03(8)	&	+1.70(30)	&	+1.60	&	1.26(12)	&	$-$1.18	&	0.018	\\
84948B$^{\ast}$	&	0.196	&	3.833(13)	&	3.70(15)	&	55	&	$-$0.82(19)	&	+1.63(30)	&	+1.75	&	1.20(12)	&	$-$1.11	&	0.020	\\
87271	&	0.149	&	3.876(13)	&	3.43(10)	&	--	&	$-$1.11(30)	&	+1.02(8)	&	+0.92	&	1.53(3)	&	$-$1.27	&	0.009	\\
102541	&	0.141	&	3.885(10)	&	4.22(16)	&	--	&	$-$0.95(20)	&	+2.34(21)	&	+2.23	&	1.01(9)	&	$-$1.30	&	0.029	\\
105058	&	0.127	&	3.889(10)	&	3.77(30)	&	140	&	$-$0.82(7)	&	+0.86(30)	&	+0.75	&	1.60(12)	&	$-$1.40	&	0.010	\\
105759	&	0.142	&	3.874(6)	&	3.65(10)	&	120	&	$-$0.92(30)	&	+1.35(21)	&	+1.25	&	1.40(8)	&	$-$1.20	&	0.015	\\
109738$^{\ast}$	&	0.144	&	3.881(8)	&	3.90(13)	&	--	&	$-$1.02(20)	&	+1.85(30)	&	+1.75	&	1.20(12)	&	$-$1.49	&	0.012	\\
110377	&	0.120	&	3.888(5)	&	3.97(14)	&	170	&	$-$0.83(20)	&	+1.96(11)	&	+1.85	&	1.16(5)	&	$-$1.45	&	0.014	\\
111604	&	0.112	&	3.890(8)	&	3.61(25)	&	180	&	$-$1.04(3)	&	+0.48(7)	&	+0.37	&	1.75(3)	&	$-$0.94	&	0.022	\\
120500	&	0.064	&	3.915(4)	&	3.86(10)	&	125	&	$-$0.73(14)	&	+0.85(34)	&	+0.70	&	1.62(13)	&	$-$1.32	&	0.014	\\
120896$^{\ast}$	&	0.166	&	3.861(5)	&	3.76(10)	&	--	&	$-$0.82(30)	&	+1.90(30)	&	+1.81	&	1.18(12)	&	$-$1.25	&	0.016	\\
125162	&	0.042	&	3.941(8)	&	4.07(9)	&	115	&	$-$1.61(24)	&	+1.71(23)	&	+1.54	&	1.28(9)	&	$-$1.64	&	0.011	\\
142703	&	0.177	&	3.861(9)	&	3.93(12)	&	100	&	$-$1.32(5)	&	+2.41(12)	&	+2.32	&	0.97(5)	&	$-$1.43	&	0.015	\\
142944$^{\ast}$	&	0.198	&	3.845(8)	&	3.19(4)	&	180	&	$-$0.91(38)	&	+0.80(30)	&	+0.69	&	1.62(12)	&	$-$0.85	&	0.016	\\
148638$^{\ast}$	&	0.106	&	3.882(13)	&	3.39(10)	&	--	&	$-$0.80(30)	&	+0.33(30)	&	+0.23	&	1.81(12)	&	$-$1.21	&	0.009	\\
153747	&	0.068	&	3.914(5)	&	3.70(24)	&	--	&	$-$0.86(20)	&	+1.24(30)	&	+1.09	&	1.46(12)	&	$-$1.31	&	0.013	\\
168740	&	0.128	&	3.883(5)	&	3.88(14)	&	145	&	$-$0.91(8)	&	+1.82(2)	&	+1.72	&	1.21(1)	&	$-$1.44	&	0.013	\\
168947$^{\ast}$	&	0.145	&	3.878(11)	&	3.67(10)	&	--	&	$-$0.74(20)	&	+1.28(30)	&	+1.18	&	1.43(12)	&	$-$1.23	&	0.014	\\
183324	&	0.032	&	3.952(10)	&	4.13(4)	&	90	&	$-$1.47(6)	&	+1.64(42)	&	+1.44	&	1.32(17)	&	$-$1.68	&	0.011	\\
191850$^{\ast}$	&	0.163	&	3.869(9)	&	3.61(10)	&	--	&	$-$0.96(30)	&	+1.50(30)	&	+1.41	&	1.34(12)	&	$-$1.13	&	0.017	\\
192640	&	0.095	&	3.900(5)	&	3.95(18)	&	80	&	$-$1.46(8)	&	+1.84(2)	&	+1.71	&	1.22(1)	&	$-$1.55	&	0.011	\\
210111	&	0.136	&	3.878(7)	&	3.84(15)	&	55	&	$-$1.04(20)	&	+1.76(15)	&	+1.66	&	1.23(6)	&	$-$1.36	&	0.014	\\
213669	&	0.155	&	3.872(8)	&	3.82(17)	&	--	&	$-$0.93(20)	&	+1.79(21)	&	+1.69	&	1.22(8)	&	$-$1.18	&	0.021	\\
221756	&	0.046	&	3.930(10)	&	3.90(3)	&	105	&	$-$0.71(3)	&	+1.16(16)	&	+1.00	&	1.50(6)	&	$-$1.36	&	0.015	\\
290799$^{\ast}$	&	0.114	&	3.889(5)	&	4.18(10)	&	70	&	$-$0.82(26)	&	+2.62(30)	&	+2.51	&	0.90(12)	&	$-$1.37	&	0.025	\\
\hline
\end{tabular}
\end{center}
\end{table*}

\begin{table*}
\caption{$\delta$ Scuti stars selected from the list by Rodriguez et
al. (2000); $\sigma$($b-y$)\,=\,$\pm$0.005\,mag; $\sigma$[Z]\,=\,$\pm$0.15\,dex. In
parenthesis are the errors in the final digits of the corresponding quantity.}
\label{dsct_l}
\begin{center}
\begin{tabular}{lcllcllllll}
\hline
\multicolumn{1}{c}{HD} & $(b-y)_{0}$ & \multicolumn{1}{c}{log\,$T_{\rm eff}$} &
\multicolumn{1}{c}{log\,$g$} & \multicolumn{1}{c}{$v$\,sin\,$i$} &  
\multicolumn{1}{c}{[Z]} & \multicolumn{1}{c}{$M_{\rm V}$} &
\multicolumn{1}{c}{$M_{\rm B}$}  & \multicolumn{1}{c}{log\,$L_{\ast}/L_{\odot}$} &
\multicolumn{1}{c}{log\,$P$} & \multicolumn{1}{c}{$Q$} \\
& [mag] & \multicolumn{1}{c}{[dex]} & \multicolumn{1}{c}{[dex]} &  
\multicolumn{1}{c}{[km\,s$^{-1}$]} & \multicolumn{1}{c}{[dex]} & \multicolumn{1}{c}{[mag]} & 
\multicolumn{1}{c}{[mag]} & \multicolumn{1}{c}{[dex]} &
& \multicolumn{1}{c}{[d]}\\
\hline
432	&	0.211	&	3.841(4)	&	3.44(7)	&	70 & +0.45	&	1.19(29)	&	1.08	&	1.47(11)	&	$-$1.00	&	0.017	\\
3112	&	0.127	&	3.883(5)	&	3.59(9)	&	80 & +0.28	&	0.54(84)	&	0.44	&	1.73(34)	&	$-$1.31	&	0.009	\\
4490	&	0.156	&	3.867(3)	&	3.55(9)	&	180 & +0.22	&	0.92(15)	&	0.83	&	1.57(6)	&	$-$0.98	&	0.019	\\
4849	&	0.168	&	3.862(2)	&	3.78(8)	&	-- & +0.52	&	1.65(30)	&	1.56	&	1.27(12)	&	$-$1.26	&	0.016	\\
7312	&	0.169	&	3.861(3)	&	3.79(6)	&	-- & +0.24	&	1.71(28)	&	1.62	&	1.25(11)	&	$-$1.38	&	0.012	\\
8511	&	0.134	&	3.880(4)	&	3.96(6)	&	190 & $-$0.06	&	2.04(1)	&	1.94	&	1.12(1)	&	$-$1.16	&	0.027	\\
8781	&	0.213	&	3.838(6)	&	3.46(6)	&	-- & $-$0.03	&	1.57(18)	&	1.46	&	1.32(7)	&	$-$0.95	&	0.020	\\
9065	&	0.200	&	3.844(6)	&	3.46(6)	&	-- & $-$0.14	&	1.72(21)	&	1.61	&	1.26(8)	&	$-$1.02	&	0.018	\\
9100	&	0.087	&	3.899(8)	&	3.53(22)	&	120 & $-$0.34	&	0.43(28)	&	0.30	&	1.78(11)	&	$-$0.87	&	0.024	\\
11522	&	0.162	&	3.861(5)	&	3.41(6)	&	120 & $-$0.04	&	0.76(13)	&	0.67	&	1.63(5)	&	$-$1.04	&	0.014	\\
15550	&	0.152	&	3.871(3)	&	3.84(6)	&	170 & +0.13	&	1.89(1)	&	1.80	&	1.18(1)	&	$-$1.17	&	0.022	\\
15634	&	0.179	&	3.858(2)	&	3.81(7)	&	140 & +0.28	&	1.57(50)	&	1.48	&	1.31(20)	&	$-$1.01	&	0.028	\\
17093	&	0.133	&	3.883(8)	&	4.04(11)	&	75 & $-$0.10	&	2.22(4)	&	2.12	&	1.05(2)	&	$-$1.45	&	0.016	\\
19279	&	0.063	&	3.913(8)	&	3.76(17)	&	285 & $-$0.16	&	1.69(61)	&	1.54	&	1.28(25)	&	$-$1.16	&	0.022	\\
23728	&	0.178	&	3.859(5)	&	3.70(16)	&	105 & $-$0.21	&	1.62(15)	&	1.53	&	1.29(6)	&	$-$1.00	&	0.025	\\
24809	&	0.119	&	3.890(10)	&	4.26(20)	&	130 & $-$0.36	&	2.51(7)	&	2.40	&	0.94(3)	&	$-$1.26	&	0.035	\\
24832	&	0.158	&	3.865(4)	&	3.69(13)	&	140 & +0.12	&	1.12(30)	&	1.03	&	1.49(12)	&	$-$0.81	&	0.035	\\
26574	&	0.196	&	3.847(4)	&	3.49(12)	&	100 & +0.62	&	1.22(38)	&	1.11	&	1.46(15)	&	$-$1.13	&	0.013	\\
27397	&	0.166	&	3.864(3)	&	3.96(5)	&	100 & +0.22	&	2.30(5)	&	2.21	&	1.02(2)	&	$-$1.26	&	0.022	\\
27459	&	0.123	&	3.884(5)	&	3.96(13)	&	75 & +0.25	&	1.92(21)	&	1.82	&	1.17(9)	&	$-$1.44	&	0.014	\\
28024	&	0.159	&	3.865(7)	&	3.40(18)	&	210 & +0.22	&	0.77(24)	&	0.68	&	1.63(9)	&	$-$0.83	&	0.022	\\
28319	&	0.093	&	3.901(9)	&	3.70(12)	&	80 & +0.16	&	0.32(78)	&	0.19	&	1.82(31)	&	$-$1.12	&	0.016	\\
28910	&	0.139	&	3.877(4)	&	3.97(5)	&	125 & +0.19	&	1.58(75)	&	1.48	&	1.31(30)	&	$-$1.17	&	0.024	\\
30780	&	0.114	&	3.887(3)	&	3.87(11)	&	150 & +0.23	&	1.41(50)	&	1.30	&	1.38(20)	&	$-$1.38	&	0.013	\\
32846	&	0.189	&	3.845(9)	&	3.37(10)	&	-- & $-$0.19	&	1.16(5)	&	1.05	&	1.48(2)	&	$-$0.87	&	0.021	\\
50018	&	0.217	&	3.836(6)	&	3.35(10)	&	135 & +0.78	&	0.49(1.03)	&	0.37	&	1.75(41)	&	$-$0.81	&	0.019	\\
57167	&	0.214	&	3.844(2)	&	3.97(3)	&	100 & +0.18	&	2.49(15)	&	2.38	&	0.95(6)	&	$-$1.33	&	0.019	\\
71496	&	0.133	&	3.878(4)	&	3.61(8)	&	130 & +0.41	&	1.18(10)	&	1.08	&	1.47(4)	&	$-$1.02	&	0.021	\\
71935	&	0.140	&	3.872(4)	&	3.67(9)	&	160 & +0.32	&	1.14(20)	&	1.04	&	1.49(8)	&	$-$1.15	&	0.016	\\
73575	&	0.137	&	3.874(5)	&	3.41(12)	&	150 & +0.32	&	0.38(21)	&	0.28	&	1.79(8)	&	$-$0.99	&	0.015	\\
74050	&	0.106	&	3.892(7)	&	3.85(16)	&	145 & +0.25	&	1.71(44)	&	1.59	&	1.26(18)	&	$-$1.24	&	0.019	\\
84999	&	0.192	&	3.851(4)	&	3.41(8)	&	110 & +0.13	&	1.09(27)	&	0.99	&	1.50(11)	&	$-$0.88	&	0.021	\\
88824	&	0.153	&	3.870(2)	&	3.83(10)	&	235 & +0.08	&	1.76(2)	&	1.67	&	1.23(1)	&	$-$0.90	&	0.039	\\
94985	&	0.088	&	3.903(9)	&	3.62(5)	&	-- & $-$0.11	&	0.82(16)	&	0.69	&	1.62(6)	&	$-$0.82	&	0.032	\\
103313	&	0.110	&	3.889(4)	&	3.67(8)	&	70 & +0.24	&	0.79(36)	&	0.68	&	1.63(14)	&	$-$1.10	&	0.017	\\
104036	&	0.086	&	3.899(7)	&	4.09(14)	&	-- & +0.01	&	1.67(40)	&	1.54	&	1.28(16)	&	$-$1.52	&	0.013	\\
107131	&	0.097	&	3.897(5)	&	4.03(20)	&	185 & $-$0.08	&	1.91(25)	&	1.78	&	1.19(10)	&	$-$1.18	&	0.029	\\
107904	&	0.224	&	3.837(6)	&	3.20(12)	&	115 & +0.68	&	0.83(15)	&	0.72	&	1.61(6)	&	$-$0.93	&	0.013	\\
109585	&	0.208	&	3.841(4)	&	3.59(10)	&	80 & +0.16	&	1.80(18)	&	1.69	&	1.22(7)	&	$-$1.09	&	0.018	\\
115308	&	0.199	&	3.846(6)	&	3.36(8)	&	75 & +0.06	&	1.16(2)	&	1.05	&	1.48(1)	&	$-$0.93	&	0.017	\\
117661	&	0.095	&	3.900(4)	&	3.95(11)	&	55 & +0.14	&	1.70(12)	&	1.57	&	1.27(5)	&	$-$1.37	&	0.016	\\
124675	&	0.111	&	3.884(6)	&	3.67(17)	&	120 & $-$0.11	&	1.02(29)	&	0.91	&	1.54(12)	&	$-$1.19	&	0.015	\\
125161	&	0.128	&	3.885(8)	&	4.10(19)	&	135 & +0.08	&	2.40(14)	&	2.29	&	0.98(5)	&	$-$1.58	&	0.013	\\
127762	&	0.112	&	3.890(3)	&	3.69(12)	&	130 & +0.02	&	0.94(2)	&	0.83	&	1.57(1)	&	$-$1.14	&	0.017	\\
127929	&	0.143	&	3.876(3)	&	3.65(11)	&	70 & +0.00	&	0.80(29)	&	0.70	&	1.62(12)	&	$-$1.06	&	0.018	\\
138918	&	0.146	&	3.870(5)	&	3.77(23)	&	85 & +0.16	&	0.19(1.13)	&	0.10	&	1.86(45)	&	$-$0.81	&	0.032	\\
143466	&	0.177	&	3.863(5)	&	3.92(15)	&	145 & +0.24	&	2.29(8)	&	2.20	&	1.02(3)	&	$-$1.12	&	0.029	\\
152569	&	0.160	&	3.864(3)	&	3.81(13)	&	195 & +0.16	&	1.83(4)	&	1.74	&	1.21(2)	&	$-$1.12	&	0.023	\\
155514	&	0.119	&	3.886(6)	&	3.90(16)	&	175 & +0.08	&	1.49(17)	&	1.38	&	1.35(7)	&	$-$1.05	&	0.029	\\
171369	&	0.159	&	3.863(3)	&	3.79(8)	&	80 & +0.05	&	1.64(20)	&	1.55	&	1.28(8)	&	$-$1.04	&	0.026	\\
176723	&	0.200	&	3.848(3)	&	3.62(9)	&	265 & +0.10	&	1.66(19)	&	1.56	&	1.27(8)	&	$-$0.87	&	0.031	\\
177392	&	0.168	&	3.861(6)	&	3.54(16)	&	140 & +0.15	&	0.96(10)	&	0.87	&	1.55(4)	&	$-$0.96	&	0.020	\\
177482	&	0.161	&	3.862(5)	&	3.45(8)	&	145 & +0.26	&	0.86(14)	&	0.77	&	1.59(6)	&	$-$1.01	&	0.016	\\
181333	&	0.138	&	3.877(4)	&	3.53(6)	&	55 & +0.38	&	0.47(47)	&	0.37	&	1.75(19)	&	$-$0.82	&	0.025	\\
182475	&	0.194	&	3.850(3)	&	3.63(14)	&	130 & +0.32	&	1.61(47)	&	1.51	&	1.30(19)	&	$-$1.11	&	0.018	\\
\hline
\end{tabular}
\end{center}
\end{table*}
\addtocounter{table}{-1}
\begin{table*}
\caption
{continued.}
\begin{center}
\begin{tabular}{lcllcllllll}
\hline
\multicolumn{1}{c}{HD} & $(b-y)_{0}$ & \multicolumn{1}{c}{log\,$T_{\rm eff}\,$} &
\multicolumn{1}{c}{log\,$g$} & \multicolumn{1}{c}{$v$\,sin\,$i$} &  
\multicolumn{1}{c}{[Z]} & \multicolumn{1}{c}{$M_{\rm V}$} &
\multicolumn{1}{c}{$M_{\rm B}$}  & \multicolumn{1}{c}{log\,$L_{\ast}/L_{\odot}$} &
\multicolumn{1}{c}{log\,$P$} & \multicolumn{1}{c}{$Q$} \\
& [mag] & \multicolumn{1}{c}{[dex]} & \multicolumn{1}{c}{[dex]} &  
\multicolumn{1}{c}{[km\,s$^{-1}$]} & \multicolumn{1}{c}{[dex]} & \multicolumn{1}{c}{[mag]} & 
\multicolumn{1}{c}{[mag]} & \multicolumn{1}{c}{[dex]} &
& \multicolumn{1}{c}{[d]}\\
\hline
185139	&	0.157	&	3.869(2)	&	3.80(7)	&	-- & +0.29	&	1.38(51)	&	1.29	&	1.39(20)	&	$-$1.19	&	0.018	\\
186786	&	0.181	&	3.856(2)	&	3.86(8)	&	-- & +0.19	&	2.09(3)	&	2.00	&	1.10(1)	&	$-$1.10	&	0.027	\\
188520	&	0.123	&	3.885(7)	&	4.05(15)	&	-- & +0.03	&	2.19(19)	&	2.08	&	1.07(8)	&	$-$1.26	&	0.025	\\
199124	&	0.167	&	3.859(3)	&	3.74(8)	&	150 & $-$0.12	&	1.89(13)	&	1.80	&	1.18(5)	&	$-$1.00	&	0.028	\\
199908	&	0.192	&	3.848(4)	&	3.42(7)	&	60 & +0.27	&	1.23(21)	&	1.13	&	1.45(8)	&	$-$1.10	&	0.013	\\
206553	&	0.171	&	3.859(4)	&	3.70(15)	&	-- & +0.20	&	1.46(2)	&	1.37	&	1.35(1)	&	$-$1.20	&	0.016	\\
208435	&	0.198	&	3.844(5)	&	3.26(8)	&	-- & +0.36	&	0.67(43)	&	0.56	&	1.68(17)	&	$-$0.83	&	0.017	\\
211336	&	0.170	&	3.862(3)	&	3.90(7)	&	90 & +0.17	&	2.12(2)	&	2.03	&	1.09(1)	&	$-$1.39	&	0.015	\\
214441	&	0.205	&	3.847(4)	&	3.55(11)	&	-- & +0.45	&	1.28(59)	&	1.18	&	1.43(23)	&	$-$0.90	&	0.024	\\
215874	&	0.163	&	3.863(4)	&	3.48(5)	&	100 & +0.23	&	0.88(20)	&	0.79	&	1.58(8)	&	$-$1.06	&	0.015	\\
217236	&	0.152	&	3.868(4)	&	3.52(8)	&	100 & +0.22	&	0.51(59)	&	0.42	&	1.73(24)	&	$-$0.90	&	0.020	\\
219891	&	0.076	&	3.902(8)	&	3.77(15)	&	165 & $-$0.02	&	0.80(6)	&	0.67	&	1.63(3)	&	$-$1.00	&	0.025	\\
220061	&	0.100	&	3.882(17)	&	3.51(2)	&	140 & +0.04	&	0.95(14)	&	0.85	&	1.56(6)	&	$-$1.27	&	0.010	\\
223781	&	0.098	&	3.897(4)	&	3.92(12)	&	165 & $-$0.20	&	1.47(8)	&	1.34	&	1.37(3)	&	$-$1.22	&	0.021	\\
\hline
\end{tabular}
\end{center}
\end{table*}

\begin{figure}
\begin{center}
\epsfxsize = 62mm
\epsffile{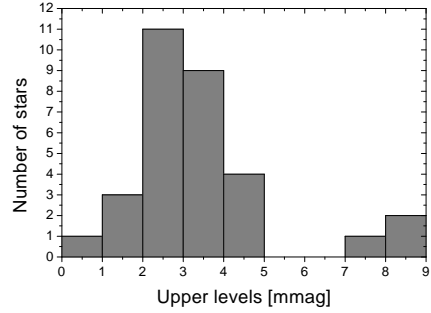}
\caption{The histogram of our upper levels for non-variability. 
See text for details.} \label{upper}
\end{center}
\end{figure}

\begin{figure}
\begin{center}
\epsfxsize = 82mm
\epsffile{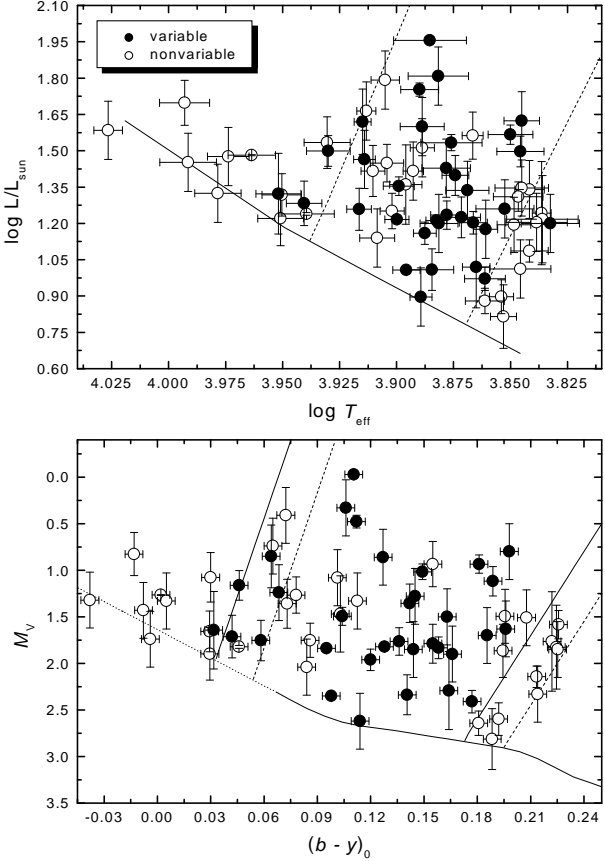}
\caption{The log\,$L_{\ast}/L_{\odot}$ versus log\,$T_{\rm eff}$ (upper panel)
and $M_{\rm V}$ versus $(b-y)_{0}$ (lower panel) diagrams for the non-variable
(open circles) and pulsating (filled circles) \LB stars. The Zero Age Main
Sequences are taken from Crawford (1979) and Claret (1995). The borders of
the classical instability strip (dotted lines) are taken from Breger (1995).
The observed borders from our sample are indicated as filled lines in the
lower panel.} \label{hrd}
\end{center}
\end{figure}

\begin{figure}
\begin{center}
\epsfxsize = 82mm
\epsffile{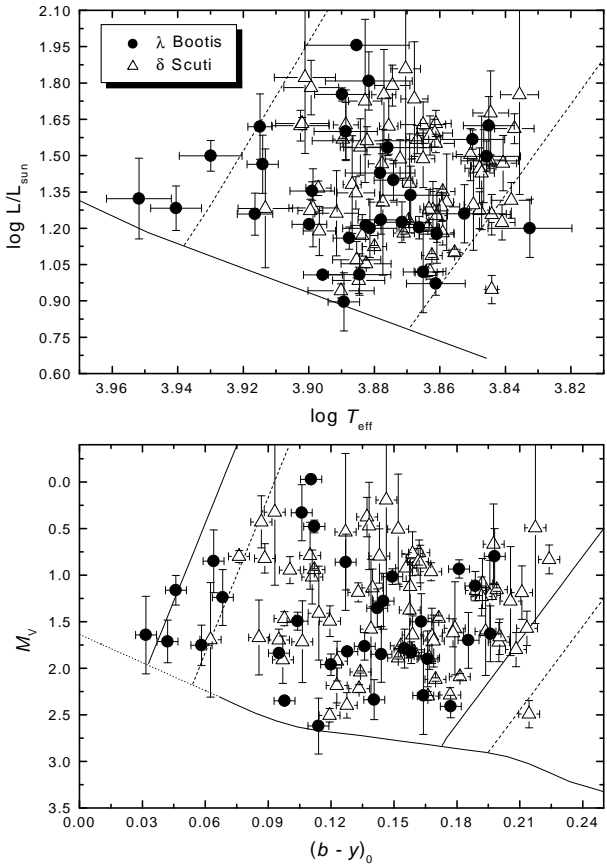}
\caption{The log\,$L_{\ast}/L_{\odot}$ versus log\,$T_{\rm eff}$ (upper panel)
and $M_{\rm V}$ versus $(b-y)_{0}$ (lower panel) diagrams for pulsating 
\LB (filled circles) and selected $\delta$ Scuti (open triangles) stars. The location
of both samples are comparable justifying our selection criteria of the $\delta$ Scuti
type stars. The lines are the same as in Fig. \ref{hrd}.} \label{hrd_all}
\end{center}
\end{figure}

\begin{figure}
\begin{center}
\epsfxsize = 75mm
\epsffile{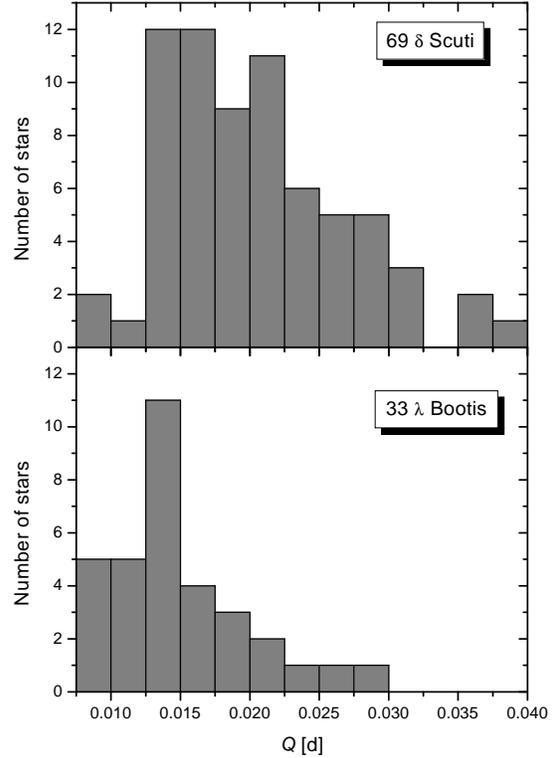}
\caption{The histograms of the pulsational constant $Q$ for
the selected $\delta$ Scuti (upper panel) and \LB (lower panel) stars.} \label{hist}
\end{center}
\end{figure}

\begin{figure}
\begin{center}
\epsfxsize = 75mm
\epsffile{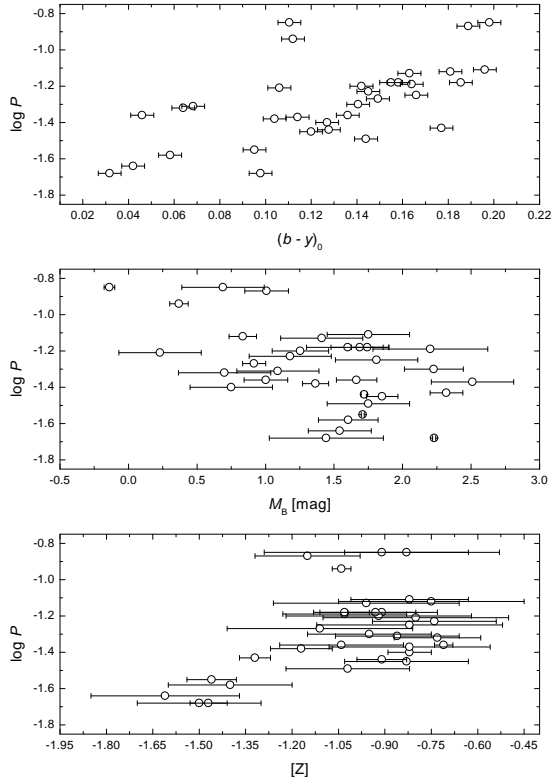}
\caption{Correlation of $(b-y)_{0}$, the absolute magnitude and metallicity with
log\,$P$ for all \LB type stars.} \label{corr}
\end{center}
\end{figure}

\begin{table}
\caption
{Nonvariable \LB stars, an asterisk denotes stars without an accurate
HIPPARCOS parallax; $\sigma$($b-y$)\,=\,$\pm$0.005\,mag. In
parenthesis are the errors in the final digits of the corresponding quantity.}
\label{constant}
\begin{center}
\begin{tabular}{lclll}
\hline
\multicolumn{1}{c}{HD} & $(b-y)_{0}$ & \multicolumn{1}{c}{log\,$T_{\rm eff}\,$} & 
\multicolumn{1}{c}{$M_{\rm V}$} &
\multicolumn{1}{c}{log\,$L_{\ast}/L_{\odot}$} \\
   & [mag] &        & \multicolumn{1}{c}{[mag]}& \\
\hline
319	&	+0.078	&	3.904(7)	&	1.27(19)	&	1.45(8)	\\
7908	&	+0.192	&	3.854(5)	&	2.60(18)	&	0.90(7)	\\
23392$^{\ast}$	&	$-$0.008	&	3.991(12)	&	1.43(30)	&	1.45(12)	\\
24472	&	+0.213	&	3.842(8)	&	2.14(11)	&	1.09(5)	\\
31295	&	+0.029	&	3.950(9)	&	1.66(22)	&	1.32(9)	\\
36726$^{\ast}$	&	$-$0.004	&	3.978(10)	&	1.74(30)	&	1.32(12)	\\
54272$^{\ast}$	&	+0.214	&	3.846(13)	&	2.33(30)	&	1.01(12)	\\
74873	&	+0.046	&	3.940(12)	&	1.82(1)	&	1.24(1)	\\
81290$^{\ast}$	&	+0.225	&	3.839(13)	&	1.85(30)	&	1.20(12)	\\
83277	&	+0.196	&	3.845(12)	&	1.49(29)	&	1.35(12)	\\
84123	&	+0.226	&	3.847(11)	&	1.58(15)	&	1.31(6)	\\
90821$^{\ast}$	&	+0.065	&	3.913(4)	&	0.74(30)	&	1.66(12)	\\
91130	&	+0.073	&	3.910(5)	&	1.36(26)	&	1.42(11)	\\
101108$^{\ast}$	&	+0.113	&	3.893(4)	&	1.33(30)	&	1.42(12)	\\
106223	&	+0.225	&	3.836(16)	&	1.83(45)	&	1.22(18)	\\
107233	&	+0.181	&	3.861(9)	&	2.64(13)	&	0.88(5)	\\
110411	&	+0.029	&	3.951(10)	&	1.90(28)	&	1.22(11)	\\
111005	&	+0.222	&	3.836(4)	&	1.76(53)	&	1.24(21)	\\
130767	&	+0.002	&	3.964(10)	&	1.27(2)	&	1.48(1)	\\
149130$^{\ast}$	&	+0.208	&	3.842(6)	&	1.51(30)	&	1.34(12)	\\
154153	&	+0.194	&	3.848(7)	&	1.86(29)	&	1.19(11)	\\
156954	&	+0.188	&	3.853(6)	&	2.81(33)	&	0.82(13)	\\
170680	&	$-$0.013	&	3.993(11)	&	0.83(23)	&	1.70(9)	\\
175445	&	+0.030	&	3.930(10)	&	1.08(27)	&	1.53(11)	\\
193256$^{\ast}$	&	+0.101	&	3.889(5)	&	1.08(30)	&	1.51(12)	\\
193281$^{\ast}$	&	+0.072	&	3.905(6)	&	0.41(30)	&	1.79(12)	\\
198160	&	+0.103	&	3.896(7)	&	1.47(41)	&	1.36(16)	\\
204041	&	+0.086	&	3.902(5)	&	1.75(18)	&	1.25(7)	\\
216847	&	+0.155	&	3.867(5)	&	0.93(24)	&	1.56(10)	\\
261904$^{\ast}$	&	+0.005	&	3.974(9)	&	1.33(30)	&	1.48(12)	\\
290492$^{\ast}$	&	+0.084	&	3.908(8)	&	2.04(30)	&	1.14(12)	\\
294253$^{\ast}$	&	$-$0.038	&	4.027(6)	&	1.32(30)	&	1.58(12)	\\
\hline
\end{tabular}
\end{center}
\end{table}

\section{Basic stellar parameters} \label{basics}

In this section we describe the calibration procedures
within various photometric systems 
and derivation of the basic stellar parameters required to analyse
the pulsational characteristics of these stars, such as the effective
temperature, surface gravity and the luminosity.

The required standard photometric colours were taken from the General
Catalogue of Photometric Data (GCPD; http://obswww.unige.ch/gcpd/) as well
as the HIPPARCOS and TYCHO databases (ESA 1997). If available, averaged
and weighted mean values were used.

The following calibrations for the individual photometric systems were
used to derive effective temperatures and surface gravities:
\begin{itemize}
\item Johnson $UBV$: Napiwotzki et al. (1993)
\item Str\"omgren $uvby\beta$: Moon \& Dworetsky (1985) and Napiwotzki et al. (1993)
\item Geneva 7-colour: Kobi \& North (1990) and K\"unzli et al. (1997)
\end{itemize}
The calibrations for the Johnson $UBV$ and
Geneva 7-colour system need an a-priori knowledge of the reddening,
which is, in general, not easy to estimate. 

Normally, the reddening for objects within the solar neighborhood is estimated
using photometric calibrations in the Str\"omgren $uvby\beta$ system
(Str\"omgren 1966; Crawford 1979; Hilditch et al. 1983). These calibrations are
not very reliable for stars with spectral types from A0 to A3 (Gerbaldi et
al. 1999), mainly because for these stars, the reddening free parameter $\beta$
is no longer a temperature indicator alone but is also sensitive to the
luminosity. From the photometry we find that two of our pulsating (HD~125162
and HD~183324) and nine constant (HD~23392, HD~31295, HD~36726, HD~74873,
HD~110411, HD~130767, HD~170680, HD~261904 and HD~294253) program stars fall
into A0 to A3 spectral region.

An independent way to derive the interstellar reddening is to use galactic
reddening maps, which are derived from open clusters as well as from galactic
field stars.  Several different models have been published in the literature
(Arenou et al. 1992, Hakkila et al. 1997). Chen et al. (1998) compared the
results from Arenou et al. (1992) and those derived from the HIPPARCOS
measurements and found an overestimation of previously published results 
from Arenou et al. (1992) for
distances less than 500\,pc. They consequently proposed a new model for
galactic latitudes of $\pm$10$\degr$, but otherwise find excellent agreement
with the model by Sandage (1972). We have used the proposed model by Chen et
al. (1998) to derive the interstellar reddening for all program stars. The
values from the calibration of the Str\"omgren $uvby\beta$ and the model by
Chen et al. (1998) are in very good agreement. To minimize possible
inconsistencies we have averaged the values from both approaches.

In Table~\ref{check}, we compare our photometrically derived effective
temperatures and surface gravities of 29 program stars from this work (TW) with those of Table
1 from Heiter et al. (2002), which contains averaged values from the literature
based on spectroscopic analyses.  The average difference for the effective
temperature is $\Delta T_{\rm eff}$\,=\,$T_{\rm eff}$(Lit.)\,$-$\,$T_{\rm
eff}$(TW)\,=\,+72(210)\,K, and for the average surface gravity $\Delta \log
g$\,=\,$\log g$(Lit.)\,$-$\,$\log g$(TW)\,=\,+0.07(24)\,dex.  We note that
there are some stars for which the spectroscopically derived values are
significantly different from the photometrically derived ones (e.g. HD~106223
and HD~107233). These cases were already extensively discussed by Heiter et
al. (2002). Although such deviating cases obviously exist, we believe that our
calibration method is consistent and therefore suitable for a statistical
analysis.

For all program stars photometrically calibrated absolute magnitudes
(assuming that all objects are single) were estimated
with an error
of $\pm$0.3\,mag. As an independent source we have taken the HIPPARCOS
parallaxes (if available) to derive absolute magnitudes using the
visual magnitude and reddening. Since we also corrected for the
Lutz-Kelker effect (Koen 1992) which is only possible for 
parallax measurements with an absolute error of
[$\sigma(\pi)/\pi$]\,$<$\,0.175 it seriously limits the useful data.
Oudmaijer et al. (1998) showed that this effect 
has to be taken into account if individual absolute magnitudes
are calculated using HIPPARCOS parallaxes. Stars without measurements
satisfying [$\sigma(\pi)/\pi$]\,$<$\,0.175 are marked with an asterisk in
Tables \ref{pulsating} and \ref{constant} (20 stars in total).
For the other 45 objects we are able to derive weighted means 
(taking the errors as weights, i.e. a larger error is a lower weight) 
for the absolute magnitude using the values from the photometric calibration
procedure and the conversion of the HIPPARCOS parallax measurements. For
the remaining 20 stars only photometrically calibrated absolute magnitudes
are available.
We then calculated luminosities (log\,L$_{\ast}$/L$_{\sun}$) using the absolute bolometric
magnitude of the Sun $M_{\rm Bol}(\sun)$\,=\,4.75\,mag (Cayrel de Strobel 1996)
and bolometric corrections taken from Drilling \& Landolt (2000).

For HD~84948B we have used the astrophysical parameters listed by Iliev et
al. (2002; Table 1). This is an evolved spectroscopic binary system which
contains two similar \LB components; Iliev et al. (2002) have taken
the binary nature into account.

Individual abundances and projected
rotational velocities for members of the \LB group were
published by Uesugi \& Fukuda (1982), Venn \& Lambert (1990), St\"{u}renburg
(1993), Abt \& Morrell (1995), Holweger \& Rentzsch-Holm (1995), Chernyshova et
al. (1998), Heiter et al. (1998), Paunzen et al. (1999a,b), Kamp et al. (2001),
Solano et al. (2001), Heiter (2002) and Andrievsky et al. (2002). The
individual values were weighted (if possible) with the errors listed in the
references and averaged.

The published
abundances do not allow an investigation of the correlation of individual
abundances of different elements (which have different diffusion properties)
with the pulsational period. It is well known that the typical abundance
pattern of \LB stars is characterized by moderate to strong underabundances of
elements heavier than C, N, O and S. To get an overall estimate of the
(surface) abundance we have applied the following method:

\begin{itemize}
\item A weighted mean for Mg, Ca, Sc, Ti, Cr and Fe was calculated and
taken as a measurement of [Z]. This should minimize measurement errors
for individual elements.
\item We determined $\Delta m_{2}$ from the Geneva 7-colour
as well as $\Delta m_{1}$ from the Str\"omgren $uvby\beta$ photometric system
(for the definition of these parameters see Golay 1974).
\item Then we correlated $\Delta m_{2}$ or $\Delta m_{1}$ with [Z] for
stars without published individual element abundances
\end{itemize}
The third step is only valid for effective temperatures cooler than 8500\,K;
otherwise the metallicity indices are no longer sensitive. This method was applied to
nine pulsating \LB type objects: HD~6870, HD~30422, HD~42503, HD~102541,
HD~109738, HD~110377, HD~153747, HD~168947 and HD~213669.

\begin{table}
\caption{Comparison of effective temperatures and surface gravities for
\LB stars from Heiter et al. (2002; Table 1; columns ``Literature'') and this work;
$\Delta T_{\rm eff}$\,=\,$T_{\rm eff}$(Lit.)\,$-$\,$T_{\rm eff}$(TW)\,=\,+72(210)\,K;
$\Delta \log g$\,=\,$\log g$(Lit.)\,$-$\,$\log g$(TW)\,=\,+0.07(24)\,dex.}
\label{check}
\begin{center}
\begin{tabular}{lccll}
\hline
& \multicolumn{2}{c}{Literature} & \multicolumn{2}{c}{This work} \\
HD & $T_{\rm eff}$ & log\,$g$ & \multicolumn{1}{c}{$T_{\rm eff}$} & 
\multicolumn{1}{c}{log\,$g$} \\
& $\pm$200\,K & $\pm$0.3\,dex \\
\hline
319 & 8100 & 3.8 & 8020(135) & 3.74(8) \\
11413 & 7900 & 3.8 & 7925(124) & 3.91(21) \\
15165 & 7200 & 3.7 & 7010(167) & 3.23(10) \\
31295 & 8800 & 4.2 & 8920(177) & 4.20(1) \\
74873 & 8900 & 4.6 & 8700(245) & 4.21(11) \\
75654 & 7250 & 3.8 & 7350(104) & 3.77(11) \\
81290 & 6780 & 3.5 & 6895(214) & 3.82(28) \\
84123 & 6800 & 3.5 & 7025(145) & 3.73(17) \\
101108 & 7900 & 4.1 & 7810(90) & 3.90(18) \\
105759 & 8000 & 4.0 & 7485(102) & 3.65(10) \\
106223 & 7000 & 4.3 & 6855(247) & 3.49(18) \\
107233 & 7000 & 3.8 & 7265(143) & 4.03(10) \\
109738 & 7575 & 3.9 & 7610(145) & 3.90(13) \\
110411 & 9100 & 4.5 & 8930(206) & 4.14(14) \\
111005 & 7410 & 3.8 & 6860(66) & 3.72(10) \\
125162 & 8650 & 4.0 & 8720(156) & 4.07(9) \\
142703 & 7100 & 3.9 & 7265(150) & 3.93(12) \\
156954 & 6990 & 4.1 & 7130(93) & 4.04(13) \\
168740 & 7700 & 3.7 & 7630(81) & 3.88(14) \\
170680 & 10000 & 4.1 & 9840(248) & 4.15(6) \\
183324 & 9300 & 4.3 & 8950(204) & 4.13(4) \\
192640 & 7960 & 4.0 & 7940(96) & 3.95(18) \\
193256 & 7800 & 3.7 & 7740(94) & 3.69(17) \\
193281 & 8070 & 3.6 & 8035(115) & 3.54(4) \\
198160 & 7900 & 4.0 & 7870(129) & 3.99(9) \\
204041 & 8100 & 4.1 & 7980(97) & 3.97(8) \\
210111 & 7530 & 3.8 & 7550(123) & 3.84(15) \\
221756 & 9010 & 4.0 & 8510(188) & 3.90(3) \\
\hline
\end{tabular}
\end{center}
\end{table}

\section{Results}

Besides two objects (HD~125889 and HD~184779), 
all members of the \LB group were photometrically
investigated. All previously published results were taken from Paunzen et
al. (1997, 1998) as well as from the references quoted in Sect. \ref{basics}.  
Of these
65 stars, 32 are presumed to be constant whereas 33 are pulsating. The upper
limits for non-variability, which are below 5\,mmag for all but three stars
(HD~31925, HD~91130 and HD~294253), are shown in Fig. \ref{upper}.

In order to investigate the pulsational characteristics of the \LB stars as a
group, we compare them with those of ``normal'' $\delta$ Scuti variables. 
The next subsection describes the compilation of
the latter sample.

\subsection{A sample of $\delta$ Scuti stars}

As a basis we have used the catalogue of Rodriguez et al. (2000). From
this sample, stars have been rejected following these criteria:
\begin{itemize}
\item Classification as Am, Ap, $\delta$ Delphini and SX Phoenicis
objects
\item $v$\,sin\,$i$\,$<$\,45\,kms$^{-1}$ (if available)
\item $\sigma(\pi)/\pi$\,$>$\,0.175
\item log $P<-1.7$ and log $P>-0.8$ 
\item Amplitude\,$>$\,0.08\,mag (if no $v$\,sin\,$i$ available)
\item without Johnson and Geneva photometry
\item $E(b-y)$\,$>$\,0.05\,mag
\end{itemize}

Such a choice is based on the characteristics of our \LB type sample and is
hoped to guarantee a comparable sample of $\delta$ Scuti type stars. In total,
69 objects remain in the sample. The basic parameters, etc. were derived in
exactly the same way as for the \LB type objects, as described in
Sect. \ref{basics}.  Table \ref{dsct_l} lists all calibrated parameters
together with the periods given by Rodriguez et al. (2000).

\subsection{Hertzsprung-Russell-diagram and the pulsational characteristics}

First of all, we have investigated the location of all \LB stars within 
the log\,$L_{\ast}/L_{\odot}$ versus log\,$T_{\rm eff}$
and $M_{\rm V}$ versus $(b-y)_{0}$ diagrams (Fig. \ref{hrd}). 
The borders of the classical instability strip are taken from Breger (1995).
There are several conclusions from this figure:
\begin{itemize}
\item The published hot and cool borders of the $\delta$ Scuti instability
strip within the $M_{\rm V}$ versus $(b-y)_{0}$ diagram do not coincide with
the observed ones for the \LB stars. The latter are bluer at the 
Zero Age Main Sequence
(ZAMS hereafter) by about
25\,mmag.  However, the borders are in accordance with the
observations within the log\,$L_{\ast}/L_{\odot}$ versus log\,$T_{\rm eff}$
diagram.
\item Taking the average of variable to non-variable objects within the classical
instability strip for both diagrams then we derive a value of at least 70\,\% pulsating
objects.
\end{itemize}
Figure \ref{hrd_all} shows the same diagrams for this sample (filled circles)
together with those of the selected $\delta$ Scuti stars (open
triangles). Besides one object (HD~57167), the cool borders are in excellent
agreement with the observations.  However, there are four hot \LB type pulsators:
HD~120500, HD~125162, HD~183324 and HD~221756. The reason for these
shifts is not yet clear. We are able to exclude measurements errors (mean
values from several references were used) and the effects of rotation (all
stars have moderate $v$\,sin\,$i$ values). In addition, taking the unreddened
colours, all four stars are still outside the hot border. Therefore it is somewhat surprising
that only one object (HD~183324) lies significantly outside the
borders within the log\,$L_{\ast}/L_{\odot}$ versus log\,$T_{\rm eff}$ diagram.

As a next step towards analyzing the
pulsational characteristics we have calculated the pulsation constants
given by
$$\log Q = -6.456 +0.5 \log g + 0.1 M_{\rm B} + \log T_{\rm eff} + \log P$$

The resulting $Q$-values are listed in Table \ref{pulsating} and in
Table~\ref{dsct_l} for our program \LB stars and for the comparison sample of
$\delta$ Scuti stars, respectively. For the \LB group, the $Q$-values range
from 0.038 to 0.033 for the fundamental radial modes and decrease to about
0.012 for the fifth radial overtone (Stellingwerf 1979; Fitch 1981). Figure
\ref{hist} (lower panel) shows a histogram of the $Q$-values for the pulsating
program stars. It seems that only a few stars, if any, pulsate in the
fundamental mode, but there is a high percentage with $Q$\,$<$\,0.020\,d (high
overtone modes).

The distribution of the $Q$-values for the \LB type stars is different from
that of the $\delta$ Scuti type sample (Fig. \ref{hist}, upper panel) at a
99.9\% level (derived from a $t$-test).

We also noticed four pulsators (HD~15165, HD~42503, HD~111604 and HD~142994)
which have considerably longer periods (log\,$P$\,$>$\,$-$0.94 corresponding to
$P$\,$<$\,8.7\,d$^{-1}$) than the rest of our sample.  They do, however, show
a similar behaviour to the remaining group members.

\subsection{The Period-Luminosity-Colour-Metallicity (PLCZ) relation}

A Period-Luminosity-Metallicity relation was found for Population\,II type
variables such as RR Lyrae and SX Phoenicis stars as well as Cepheids (Nemec et
al. 1994). These objects pulsate in the radial fundamental, first and second
overtone modes. The dependence of the pulsational period on the metallicity is
purely evolutionary, i.e. older objects exhibit a lower overall abundance and a
different pulsational period. The Period-Luminosity-Metallicity relation serves as a
distance indicator widely used for extragalactic objects.

The situation for \LB stars is very different. All evidence
indicates that we find only peculiar
surface abundances whereas the overall abundance of the stars is solar,
i.e. these stars are true Population\,I objects. 
The conclusion that \LB stars are true Population\,I objects is based on
their galactic space motions (Faraggiana \& Bonifacio 1999) combined
with their location in the Hertzsprung-Russell-diagram (Fig. \ref{hrd}).
With the exception of the SX Phe stars, Population\,II type objects are
located at much higher absolute magnitudes and thus luminosities than
found for the \LB group. However, the space motions of SX Phe stars are
inconsistent with Population\,I, which facilitates an easy separation
from \LB stars.

To examine the presence of a PLCZ relation, the following basic approach
was chosen:
$$\log P = a + b(b-y)_{0} +c(M_{\rm B}) + d{\rm [Z]} $$ The [Z]-values range
from $-$0.71\,dex to $-$1.61\,dex for the \LB type sample whereas the $\delta$
Scuti type stars have values from $-$0.36\,dex to +0.78\,dex compared to the
Sun. The coefficients for the PLCZ relation were determined simultaneously,
applying a multiregression analysis (Christensen 1996). This takes into account
the individual errors of the bolometric absolute magnitude and metallicity as
weights, whereas the errors for the period and colour were assumed to be
constant for all stars. The solution was determined using a least-squares fit
and a maximum-likelihood method.  Both give consistent results, as summarized
in Table \ref{plcz}. Figure \ref{corr} shows the individual correlations.
The correlations of the bolometric magnitude and color with the pulsational period
are compatible with those found for the $\delta$ Scuti stars.

\begin{table}
\caption
{Estimates for log\,$P$\,=$a$\,+\,$b(b-y)_{0}$\,+\,$c(M_{\rm B})$\,+\,$d{\rm [Z]}$
for all pulsating program stars (left upper column), for stars with
[Z]\,$>$\,$-$1.3 excluding HD~30422, HD~35242,
HD~125162, HD~142703, HD~183324 and HD~192640 (right upper column)
as well as for the selected
$\delta$ Scuti stars (lower column); $F$ denotes the significance level of the test 
for a zero
hypothesis; in brackets are the standard errors of the estimates.}
\label{plcz}
\begin{center}
\begin{tabular}{llr|lr}
\hline
coeff. & \multicolumn{1}{c}{value} & \multicolumn{1}{c|}{$F$} & 
\multicolumn{1}{c}{value} & \multicolumn{1}{c}{$F$} \\
& &
\multicolumn{1}{c|}{[\%]} & & \multicolumn{1}{c}{[\%]} \\
\hline
\LB \\
$a$ & $-$1.25(9) & $<$0.01 & $-$1.41(9) & $<$0.01 \\
$b(b-y)_{0}$ & +3.01(41) & $<$0.01 & +2.71(46) & $<$0.01 \\
$c(M_{\rm B})$ & $-$0.20(3) & $<$0.01 & $-$0.19(3) & $<$0.01 \\
$d{\rm [Z]}$ & +0.14(6) & 13.20 & $-$0.06(17) & 74.49 \\
\hline
$\delta$ Scuti \\
$a$ & $-$1.17(6) & $<$0.01 \\
$b(b-y)_{0}$ & +2.71(43) & $<$0.01 \\
$c(M_{\rm B})$ & $-$0.23(3) & $<$0.01 \\
$d{\rm [Z]}$ & $-$0.28(7) & $<$0.01 \\
\hline
\end{tabular}
\end{center}
\end{table}

\begin{figure}[t]
\begin{center}
\epsfxsize = 70mm
\epsffile{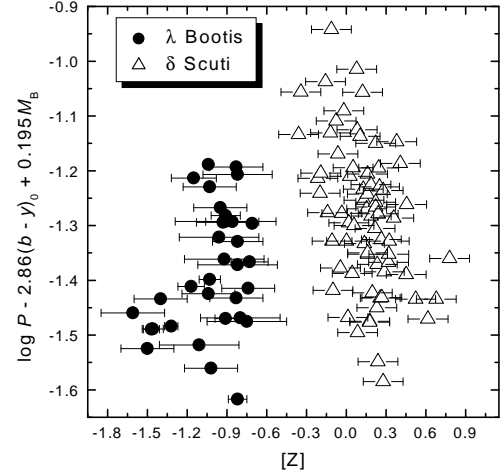}
\caption{Correlation of [Z] for the \LB (filled circles) and
selected $\delta$ Scuti type (open triangles) stars.} \label{z_all}
\end{center}
\end{figure}

To investigate whether the [Z] term is indeed significant, a plot
[log\,$P$\,$-$\,$2.86(b-y)_{0}$\,+\,0.195$(M_{\rm B})$] versus [Z] was drawn
(Fig. \ref{z_all}). The coefficients for $(b-y)_{0}$ and $M_{\rm B}$ are the
mean values from Table \ref{plcz} and are consistent within the errors for both
the $\delta$~Scuti and \LB samples. Figure \ref{z_all} shows that both samples
exhibit a trend with [Z] (with an offset of about 1\,dex). 
Whereas the [Z] term is statistically significant
for the selected sample of $\delta$ Scuti stars, it is caused by only six \LB
stars with strong underabundances (HD~30422, HD~35242, HD~125162, HD~142703,
HD~183324 and HD~192640) and vanishes after excluding them. 
We have also tested the sample for possible correlations of the [Z] term by excluding other
data points. We find no other
selection criteria of objects by means of a physical explanation, only ``suitable''
discarding would yield a clear correlation. This implies that
the peculiar abundances do not affect the pulsational period for the group of
\LB type stars. However, we find within the errors no difference of the PLC
relation for the \LB and $\delta$ Scuti type stars.

\section{Conclusions}

We have investigated the pulsational characteristics of a group of \LB
stars and compared it to a sample of $\delta$ Scuti pulsators. The latter
was chosen such that it matches our program stars within the global
astrophysical parameters. The following properties of the \LB stars are
different from those of the $\delta$ Scuti pulsators:
\begin{itemize}
\item At least 70\% of all \LB types stars inside the classical instability
strip pulsate
\item Only a maximum of two stars pulsate in the fundamental mode but 
there is a high percentage with $Q$\,$<$\,0.020\,d (high overtone modes) 
\item The instability strip of the \LB stars at the ZAMS is 25\,mmag bluer in
$(b-y)_0$ than that of the $\delta$ Scuti stars.
\end{itemize} 
We find no clear evidence for a significant term for a [Z] correlation with the
period, luminosity and colour but the PLC relation is within the errors
identical with that of the $\delta$ Scuti type stars. We note that for all but one of the
investigated pulsators, high-degree nonradial modes were detected
spectroscopically (Bohlender et al. 1999), which represents excellent agreement
with our work. The spectral variability of the \LB stars is
very similar to that seen in rapidly rotating $\delta$ Scuti stars
(Kennelly et al. 1992).

\begin{acknowledgements}
This work benefitted from the Fonds zur F\"orderung der wissenschaftlichen
Forschung, project {\em P14984}. ERC would like to thank D.~Romero,
E.~Colmenero and S.~Potter for their support.  Use was made of the SIMBAD
database, operated at CDS, Strasbourg, France and the GCPD database, operated
at the Institute of Astronomy of the University of Lausanne. We are also indebted
to the committees of the SAAO, Siding Spring and Fairborn Observatory who granted
observing time.
\end{acknowledgements}

\end{document}